\documentclass[final,5p,times,twocolumn]{elsarticle}

\usepackage{fancyhdr}
\fancypagestyle{pprintTitle}{
  \fancyhf{}
  \fancyhead[R]{CTPU-PTC-26-12, TUM-EFT 211/26}
   
  \fancyfoot[L]{}
}

\usepackage{amssymb}
\usepackage{amsmath}

\usepackage{xcolor}
\usepackage{hyperref}
\usepackage{url}

\newcommand\beq{\begin{eqnarray}}
\newcommand\eeq{\end{eqnarray}}

\newcommand\Fig[1]{Fig.~\ref{fig:#1}}
\newcommand\Eq[1]{Eq.~\eqref{eq:#1}}

\begin{document}


\begin{frontmatter}



\title{Lattice studies of chimera baryons in Sp(4) gauge theory} 


\author[1]{Jong-Wan Lee\fnref{fn}} 
\ead{j.w.lee@ibs.re.kr}
\author[2,3]{Ed Bennett}
\author[4]{Luigi Del Debbio}
\author[3,5]{Niccol\`{o} Forzano}
\author[4,6]{Ryan C. Hill}
\author[7,8]{Deog Ki Hong}
\author[9]{Ho Hsiao}
\author[10,11]{C.-J. David Lin}
\author[12]{Biagio Lucini}
\author[13]{Alessandro Lupo}
\author[3,5]{Maurizio Piai}
\author[14]{Davide Vadacchino}
\author[3,5,15]{Fabian Zierler}

\fntext[fn]{Speaker}

\affiliation[1]{organization={Center for Theoretical Physics of the Universe},
            addressline={Institute for Basic Science}, 
            city={Daejeon},
            postcode={34126}, 
            country={Korea}}

\affiliation[2]{organization={Swansea Academy of Advanced Computing, Swansea University (Bay Campus)},
            addressline={Fabian Way}, 
            city={Swansea},
            postcode={SA1 8EN}, 
            country={United Kingdom}}

\affiliation[3]{organization={Centre for Quantum Fields and Gravity, Faculty of Science and Engineering, Swansea University},
            addressline={Singleton Park}, 
            city={Swansea},
            postcode={SA2 8PP}, 
            country={United Kingdom}}

\affiliation[4]{organization={Higgs Centre for Theoretical Physics, School of Physics and Astronomy, The University of Edinburgh},
            addressline={Peter Guthrie Tait Road}, 
            city={Edinburgh},
            postcode={EH9 3FD}, 
            country={United Kingdom}}

\affiliation[5]{organization={Department of Physics, Faculty of Science and Engineering, Swansea University},
            addressline={Singleton Park}, 
            city={Swansea},
            postcode={SA2 8PP}, 
            country={United Kingdom}}

\affiliation[6]{organization={School of Physics and Astronomy, The University of Edinburgh},
            addressline={}, 
            city={Edinburgh},
            postcode={EH9 3FD}, 
            country={United Kingdom}}

\affiliation[7]{organization={Department of Physics, Pusan National University},
            addressline={}, 
            city={Busan},
            postcode={46241}, 
            country={Korea}}

\affiliation[8]{organization={Extreme Physics Institute, Pusan National University},
            addressline={}, 
            city={Busan},
            postcode={46241}, 
            country={Korea}}

\affiliation[9]{organization={Center for Computational Sciences, University of Tsukuba},
            addressline={1-1-1 Tennodai}, 
            city={Tsukuba},
            postcode={305-8577}, 
            state={Ibaraki},
            country={Japan}}

\affiliation[10]{organization={Institute of Physics, National Yang Ming Chiao Tung University},
            addressline={1001 Ta-Hsueh Road}, 
            city={Hsinchu},
            postcode={30010}, 
            country={Taiwan}}

\affiliation[11]{organization={Centre for High Energy Physics, Chung-Yuan Christian University},
            addressline={Chung-Li}, 
            city={},
            postcode={32023}, 
            country={Taiwan}}

\affiliation[12]{organization={School of Mathematical Sciences, Queen Mary University of London},
            addressline={Mile End Road}, 
            city={E1 4NS},
            postcode={London}, 
            country={United Kingdom}}

\affiliation[13]{organization={Aix Marseille Univ, Universit\'{e} de Toulon},
            addressline={CNRS, CPT}, 
            city={Marseille},
            postcode={}, 
            country={France}}
            
\affiliation[14]{organization={Centre for Mathematical Sciences, University of Plymouth},
            addressline={}, 
            city={Plymouth},
            postcode={PL4 8AA}, 
            country={United Kingdom}}

\affiliation[15]{organization={Technical University of Munich, TUM School of Natural Sciences, Physics Department},
            addressline={James-Franck-Str. 1}, 
            city={Garching},
            postcode={85748}, 
            country={Germany}}

\begin{abstract}
We  study chimera baryons, fermion bound states composed of two (hyper)quarks transforming in the fundamental and one in the antisymmetric representation of a non-Abelian gauge group. While in QCD they coincide  with ordinary baryons, in composite Higgs models (CHMs) with top partial compositeness, spin-1/2 chimera baryons serve as partners of the top quark and are responsible for its large mass. We perform non-perturbative lattice calculations of the low-lying spectrum of the chimera baryons, in a specific realization of CHMs based on a Sp(4) gauge theory. In the quenched approximation, we present the numerical results in the continuum and massless limits. Then, for dynamical fermions, we measure the spectrum and matrix elements by employing a newly developed spectral density analysis for several choices of the lattice parameters. 
\end{abstract}






\end{frontmatter}



\section{Introduction}
\label{intro}

In quantum chromodynamics (QCD), baryons are bound states composed of three quarks transforming in the fundamental representation of the ${\rm SU}(3)$ gauge group. 
In ${\rm SU}(3)$, the fundamental quark and the antisymmetric antiquark coincide, and 
thus baryons and states composed of two fundamental quarks and one antisymmetric antiquark are indistinguishable. 
This is not the case for non-Abelian gauge groups other than ${\rm SU}(3)$.
In ${\rm SU}(N _c)$, baryons are totally antisymmetric combinations of $N_c$ fundamental quarks, and in the 't Hooft large-$N_c$ limit have ${\cal O}(N_c)$ masses, compared to mesons,  quark-antiquark bound states with ${\cal O}(1)$ masses~\cite{tHooft:1973alw,Witten:1979kh}---except for the pseudo-Nambu-Goldstone Bosons (pNGB). Chimera baryons, aforementioned bound states composed of two fundamental quaks and one antisymmetric antiquark,  have  ${\cal O}(1)$ masses, and thus can appear in the low-lying spectrum at large-$N_c$, 
along with the mesons~\cite{Corrigan:1979xf}. 
For even values of $N_c$, furthermore, baryons are bosons, while chimera baryons are always fermions.

In symplectic gauge groups, ${\rm Sp}(N_c=2N)$, baryons composed of even numbers of fundamental fermions are always bosons, that decay into multiplets that, because of the symmetry enhancement and the pseudoreal nature of the fundamental representation, comprise both diquarks and mesons~\cite{Bennett:2023mhh}. 
By contrast, chimera baryons are fermion bound states with ${\cal O}(1)$ masses regardless of $N_c$. 
With the additional assumption of large-$N_c$ universality~\cite{Lovelace:1982hz}, 
therefore, chimera baryons in ${\rm Sp}(N_c)$ gauge theories provide an interesting alternative  approximation to QCD baryons---with caveats about flavour symmetries~\cite{Cherman:2009fh}. 

Interest in ${\rm Sp}(2N)$ theories and their chimera baryons has also phenomenological motivation, in extensions of the Standard Model (SM). The little hierarchy problem, the tension between current bounds on new
 physics scales and observed Higgs mass, can be addressed by interpreting the light Higgs fields as pNGBs arising in novel, strongly-coupled (hypercolour) theories, exploiting an analogy with the pions in QCD (see, e.g., Ref.~\cite{Panico:2015jxa} and references therein). 
The large mass of the top quark can be accommodated via partial compositeness, as arising from the mixing between SM quarks and fermion composite states of the new strong sector. 
Compelling candidates for such states are the spin-1/2 chimera baryons 
in the ${\rm Sp}(4)$ gauge theory coupled to two fundamental and three antisymmetric hyperquarks~\cite{Ferretti:2013kya,Barnard:2013zea}. The extended global flavour symmetry, ${\rm SU}(4)\times {\rm SU}(6)$,  is broken to  its ${\rm Sp}(4)\times {\rm SO}(6)$ subgroup, in the presence of finite hyperquark masses and condensates.

In this contribution, we present numerical results, obtained from non-perturbative lattice calculations with a focus on Sp($4$) gauge group, for two sets of physical quantities central to phenomenological studies of these theories: the masses and matrix elements of the lightest chimera baryons.  Full results are found in recent publications~\cite{Bennett:2023mhh,TELOS:2025ash} (see also~\cite{Bennett:2022yfa} and the review~\cite{Bennett:2023wjw}).
For the quenched approximation we review the mass spectrum extrapolated to continuum and massless limits, 
while for dynamical fermions we demonstrate the reach of a newly proposed spectral density method~\cite{Hansen:2019idp} on selected ensembles.

\section{Lattice action and observables}
\label{sec2}

We define the discretized Euclidean lattice action on a hypercubic lattice of size $T \times L^3=a^4 N_t \times N_s^3$, where $a$ is the lattice spacing.  We employ the standard plaquette gauge action, in which the gauge links are elements of ${\rm Sp}(4)$, and the Wilson-Dirac hyperquark fields, $Q$ and $\Psi$, transform in the fundamental, $({\rm f})$, and antisymmetric, $({\rm as}),$ representations, respectively~\cite{Bennett:2022yfa}. 
The action depends on three lattice parameters: the coupling, $\beta=8/g^2$, with $g$ the gauge coupling, 
and two masses, $m_0^{\rm f}$ and $m_0^{\rm as}$, for the $Q$ and $\Psi$ hyperquarks.  
We note that the Wilson lattice formulation breaks the global symmetry in the same way as the corresponding continuum theory. As our scale setting procedure~\cite{Bennett:2022ftz}, we employ the gradient flow method and present all the dimensionful quantities in units of the gradient flow scale, $w_0$, and introduce the hatted notation, $\hat{m} = w_0 m$.

\begin{figure}[t]
\begin{center}
\centering
\includegraphics[width=1.0\columnwidth]{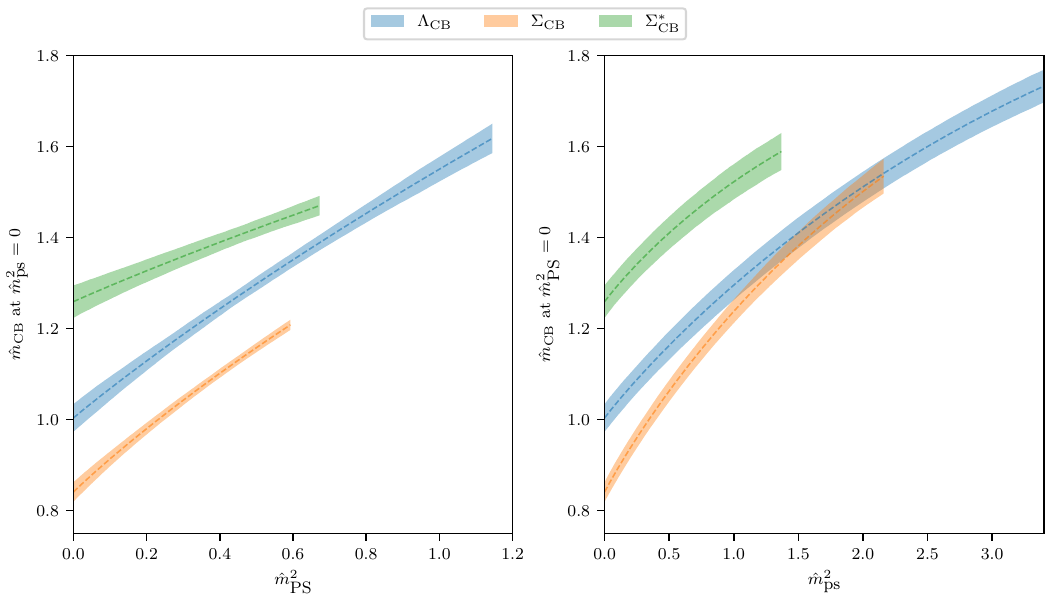}
\caption{
Continuum-extrapolated masses taken from Ref.~\cite{Bennett:2023mhh},  for (parity-even) chimera baryons, $\Lambda_{\rm CB}$, $\Sigma_{\rm CB}$, and $\Sigma^*_{\rm CB}$ in the quenched Sp$(4)$ gauge theory, as a function of the mass squared,
$\hat{m}_{\rm PS}^2$ (left) and $\hat{m}_{\rm ps}^2$ (right), of mesons  composed of $({\rm f})$- and $({\rm as})$-type hyperquark constituents.
}
\label{fig:fig_cb_fit}       
\end{center}
\end{figure}

The main observables of interest are the masses and matrix elements of chimera baryons composed of two $({\rm f})$- and one $({\rm as})$-type
hyperquark.
We extract these quantities from the lattice measurements of the $2$-point correlation functions for chimera baryons, 
$C_{CB}(t) \equiv \langle {\cal O}_{\rm CB}(t)\, {\cal O}_{\rm CB}^\dagger (0) \rangle$, by defining
\beq
{\cal O}_{\rm CB,R}=(Q^a C\Gamma Q^b)\Omega^{ac} \Omega^{bc} \Psi^{cb},
\label{eq:ops}
\eeq
where $a,b,c,d=1,\cdots,4$ denote hypercolour indexes,
$\Omega$ is the $4\times 4$ symplectic matrix, and $C$ is the charge-conjugation matrix. 
The irreducible representation, $R$, of the global symmetry  group is
$R=5$ and $10$, for the gamma structures $\Gamma=\gamma^5$ and $\gamma^i$, with $i=1,2,3$, respectively. 
While the former has the total spin quantum number $J=1/2$, the latter can be decomposed into $J=1/2$ and $3/2$ states.  Borrowing the QCD notation for baryons with strangeness $S=1$, 
we denote the corresponding chimera baryons by $\Lambda_{\rm CB}$, $\Sigma_{\rm CB}$, and $\Sigma_{\rm CB}^*$.  We perform spin projections for ${\cal O}_{{\rm CB},10}$ as 
$C_{\Sigma_{\rm CB}}(t)=P^{1/2} C_{{\rm CB},10}(t)$ and 
$C_{\Sigma^*_{\rm CB}}(t)=P^{3/2} C_{{\rm CB},10}(t)$, 
with the projectors $P^{1/2}=\frac{1}{3}\gamma^i\gamma^j$ and $P^{3/2}=\delta^{ij}-\frac{1}{3}\gamma^i\gamma^j$, respectively~\cite{Bennett:2023mhh}. 
Parity even (+) and odd (-) states are obtained 
using the parity operator $P_\pm \equiv(1\pm \gamma^0)/2$.

\section{Quenched fermions}
\label{sec3}

We performed a study of the quenched approximation, generating gauge ensembles with five different values of the lattice coupling,  $\beta=7.62,\,7.7,\,7.85,\,8.0,\,8.2$, on lattices with extent $60\times 48^3$, except for the coarsest one with extent $48\times 24^3$~\cite{Bennett:2023mhh,Bennett:2023qwx}. We use the heat-bath update algorithm,  along with over-relaxation techniques, implemented on the HiRep code \cite{DelDebbio:2008zf} adapted to ${\rm Sp}(2N)$ gauge theories~\cite{Bennett:2017kga,Bennett:2019cxd,Bennett:2020qtj}.   

\begin{figure}[t]
\centering
\includegraphics[width=1.0\columnwidth]{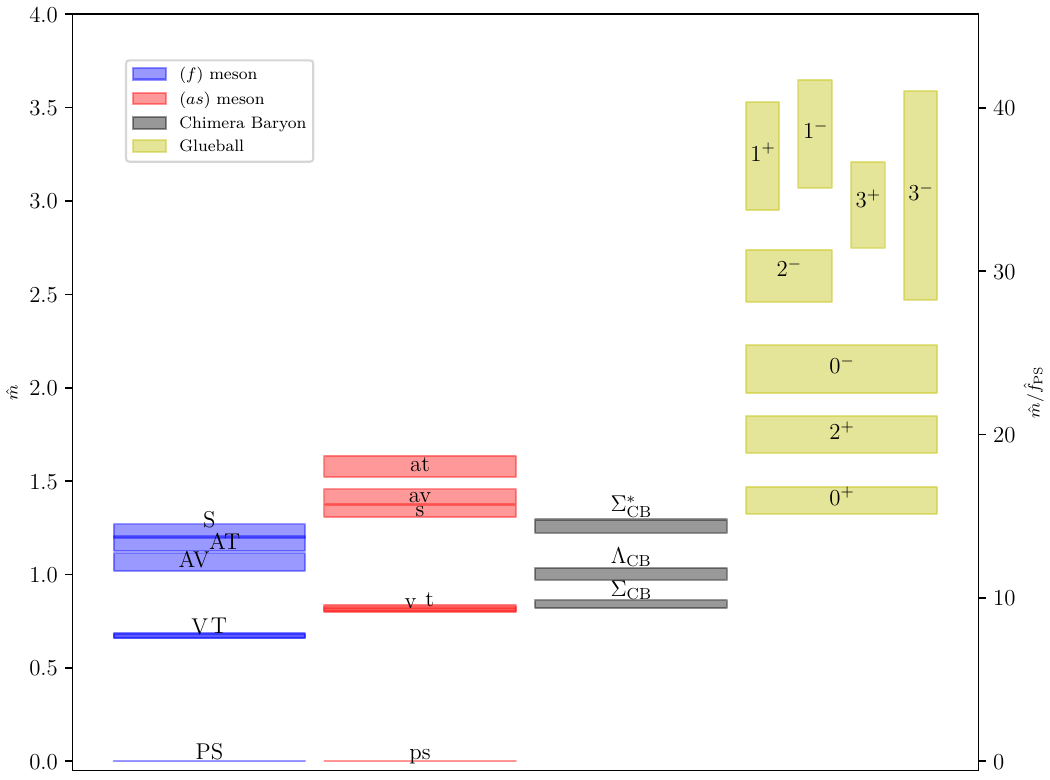}
\caption{
Summary plot for the mass spectrum in the quenched approximation for the ${\rm Sp}(4)$ gauge theory, extrapolated to the continuum and massless-hyperquark limits~\cite{Bennett:2023mhh}.
We denote mesons composed of  $(f)$-type ($(as)$-type) hyperquarks as
PS (ps), V (v), T (t), AV (av), AT (at), S (s), according to the gamma-matrix structure of the corresponding hyperquark bilinear operator~ \cite{Bennett:2019cxd}. We also show the glueball states, identified by their spin and parity quantum numbers, $J^P$~\cite{Bennett:2020qtj}. 
}%
\label{fig:fig_qc_spec}      
\end{figure}

We compute $2$-point correlation functions of chimera baryons, projected onto designated 
spin and parity states, and determine their masses by fitting the results to a single exponential function 
over a plateau region appearing at large Euclidean time, restricting attention to the lightest states only.
We improved the signal by applying Wuppertal and APE smearing techniques---details of the implementation and parameter choices are found in Ref.~\cite{Bennett:2023mhh}. 
Over the range of pNGB mass, $\hat{m}_{\rm PS} \in [0.28, 1.03]$ and $\hat{m}_{\rm ps} \in [0.35,1.84]$ for $(f)$-type and $(as)$-type hyperquarks, 
we conduct about $150$ measurements. 

We then carry out the continuum and massless extrapolations by fitting the lattice results  
to fitting ansatzes inspired by the baryon chiral perturbation theory for QCD, by including correction terms 
that are quadratic and cubic in $\hat{m}_{\rm PS}$ and/or $\hat{m}_{\rm ps}$, 
and linear in $\hat{a}$---see Ref.~\cite{Bennett:2023mhh} for the full procedure of data selection, 
as well as the best fit results. We show the dependence of $\hat{m}_{\Lambda_{\rm CB}}$, 
$\hat{m}_{\Sigma_{\rm CB}}$, and $\hat{m}_{\Sigma^*_{\rm CB}}$, on the pNGB mass squared $\hat{m}_{\rm PS}^2(m_{\rm ps}^2)$, after taking $\hat{m}_{\rm ps}(\hat{m}_{\rm PS})\rightarrow0$ and the continuum extrapolation, in \Fig{fig_cb_fit}. 
In the range of hyperquark masses considered,  we find that $\Lambda_{\rm CB}$ is heavier than $\Sigma_{\rm CB}$, but lighter than $\Sigma^*_{\rm CB}$.  In \Fig{fig_qc_spec}, we further compare the masses of chimera baryons
with those of mesons and glueballs in the accessible channels, characterising them  with spin and parity quantum numbers. The lightest chimera baryon, $\Sigma_{\rm CB}$, has a mass 
similar to that of the vector meson composed $(as)$-type hyperquarks.

\section{Dynamical fermions}
\label{sec4}

\begin{figure}[t]
\centering
\includegraphics[width=1.0\columnwidth]{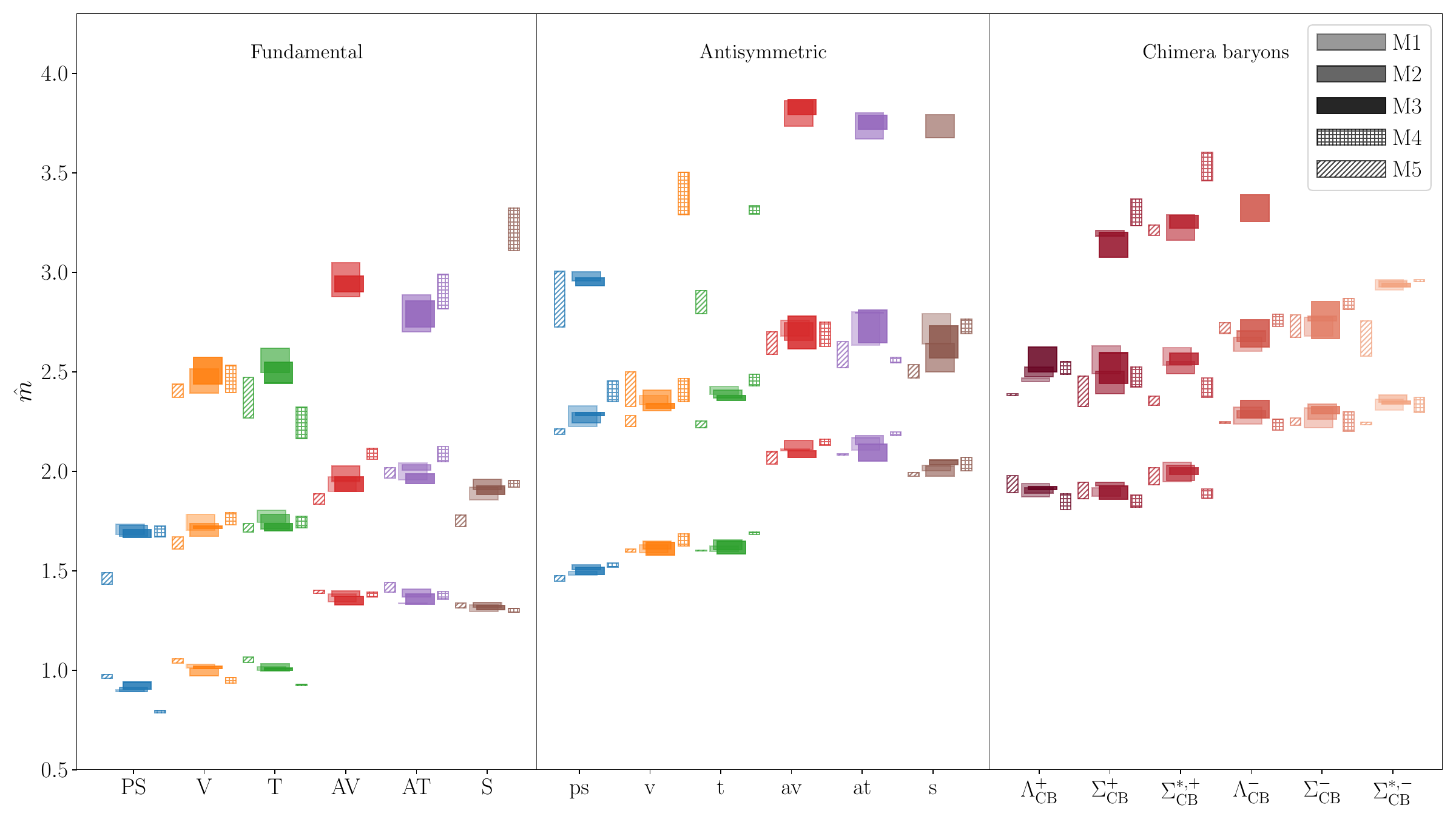}
\caption{
Summary plot for the mass spectrum of the ${\rm Sp}(4)$ gauge theory coupled to 
dynamical hyperquarks in the fundamental and antisymmetric representations~\cite{TELOS:2025ash}. 
 The ensembles M1 to M5 are characterised by different choices of lattice parameters, as discussed in the main text. 
The identifications of mesons and chimera baryons use the same conventions as in \Fig{fig_qc_spec}, 
with additional notation for parity-even (+) and parity-odd (-) chimera baryons. 
}
\label{fig:fig_dyn_spec}       
\end{figure}

For calculations with dynamical fermions, we generated five ensembles with fixed lattice coupling and mass of the $({\rm as})$-type hyperquarks, $\beta=6.5$ and $am_0^{\rm as}=-1.01$. We used the
 (R)HMC algorithms implemented in the GRID code with an extension for ${\rm Sp}(N_c)$ gauge theories~\cite{Boyle:2015tjk,Bennett:2023gbe}. 
We label the  ensembles by the $({\rm f})$-type hyperquark mass and lattice volume, as $(am_0^{\rm f}, N_t, N_s)$, 
 ${\rm M1}=(-0.71, 48, 20)$, ${\rm M2}=(-0.71, 64, 20)$, ${\rm M3}=(-0.71, 96, 20)$, ${\rm M4}=(-0.7, 63, 20)$, and ${\rm M5}=(-0.72, 64, 32)$~\cite{TELOS:2025ash,Bennett:2024cqv}.

We compute the $2$-point correlation functions of the chimera baryons by applying APE and Wuppertal smearing to the operators in \Eq{ops} (see also Refs.~\cite{Bennett:2023rsl,Bennett:2024wda} for the use of smearing to measure flavour-singlet mesons). 
We then reconstruct the spectral density from the measured correlation functions on a finite lattice by adopting the Hansen-Lupo-Tantalo method \cite{Hansen:2019idp}. 
The key idea of this approach is to introduce the smeared spectral density, $\rho_\sigma(\omega)\equiv \int_{E_{\rm min}}^\infty dE\, \Delta_\sigma(E-\omega) \rho(E)$, 
with a choice of smearing kernel, $\Delta_\sigma$, and approximate it with $\hat{\rho}_\sigma$, as a finite sum of the measured correlation functions. 
Following the procedure developed for meson operators \cite{Bennett:2024cqv}, 
we consider both Gaussian and Cauchy kernels to estimate the systematic errors 
associated with the choice of kernels. 
We extract the energy spectra by fitting the results of $\hat{\rho}_\sigma$ to the fitting function, $f_\sigma^{(k)}(E) = \sum_{n=0}^{k-1} {\cal A}_n \Delta_\sigma (E-E_n)$. 
The number of energy levels, $k$, is determined a posteriori, by examining the stability of the fits. 
The coefficient ${\cal A}_n$ encodes the (bare) matrix elements between the vacuum and chimera baryon states, ${\cal A}_n=|K^0_{\rm CB}|^2$.  The resulting
overlap factors receive multiplicative renormalisations, as $K_{\Lambda(\Sigma)} = Z_{{\rm CB},\gamma_5(\gamma_i)} |K^0_{\Lambda(\Sigma)}|$. 
We determine the renormalisation factors through one-loop perturbative matching with  tadpole improvement in the $\overline{\rm MS}$ scheme
\begin{equation}
Z_{{\rm CB},\gamma_5}=1+\frac{g^2}{16\pi^2\langle P \rangle}\left[\left(C^f+\frac{1}{2}C^{as}\right)\Delta_{\Sigma_1}+\Delta_{\rm CB}[\gamma_5] \right]\,,
\end{equation}
and analogous expression for $\gamma_5\rightarrow\gamma_i$.
Here, $\langle P \rangle$ is the average plaquette value and $C^R$ the quadratic Casimir for the representation $R$. 
Besides $\Delta_{\Sigma_1}=-12.82$, we find that $\Delta_{\rm CB}[\gamma_5]=-26.67$, and $\Delta_{\rm CB}[\gamma_i]=18.12$, associated with the vertex diagrams~\cite{TELOS:2025ash}.

\begin{figure}[t]
\centering
\includegraphics[width=1.0\columnwidth]{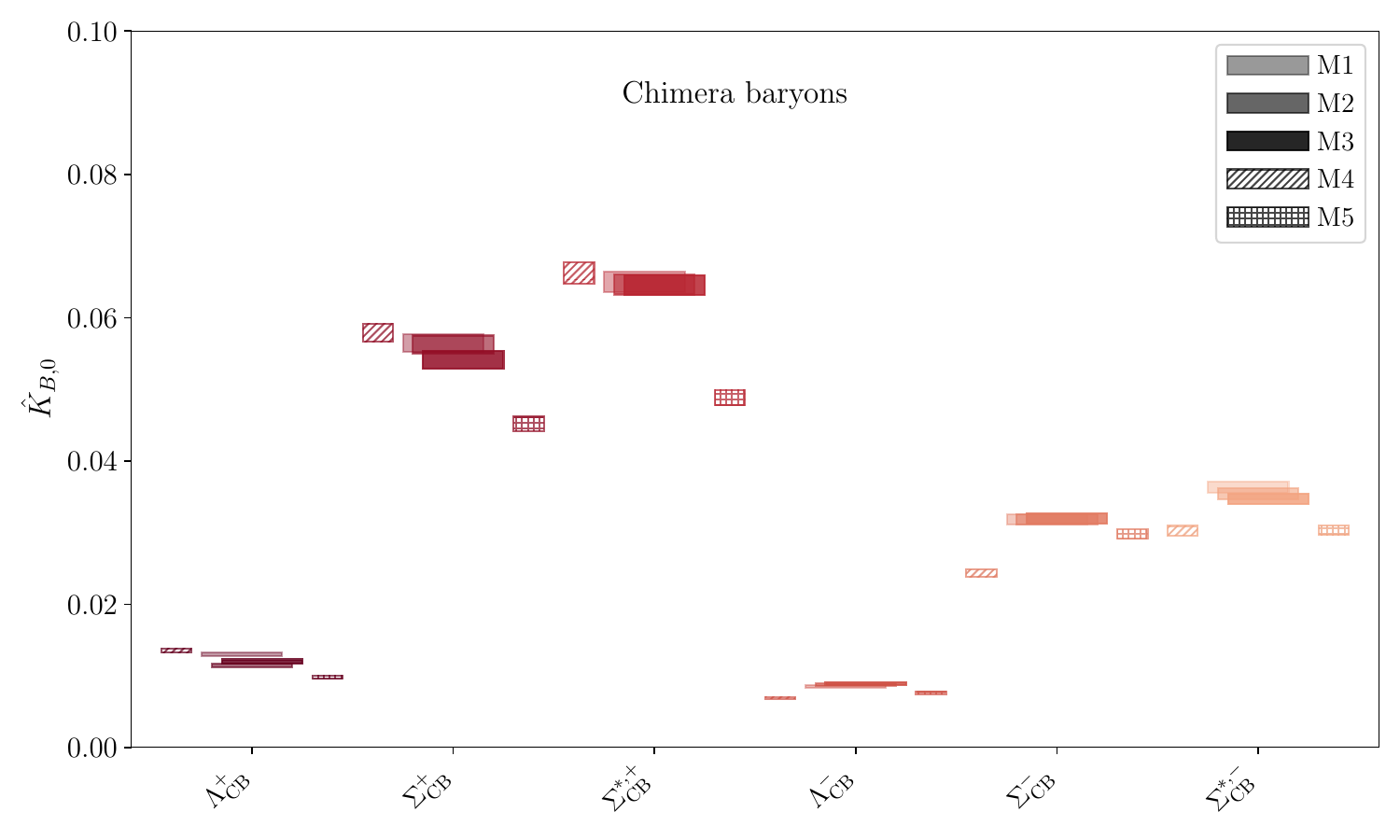}
\caption{Renormalised overlap factors of chimera baryons in the ${\rm Sp}(4)$ gauge theory coupled to dynamical hyperquarks in the fundamental and antisymmetric representations~\cite{TELOS:2025ash}.
}%
\label{fig:fig_cb_mat_el}       
\end{figure}

We summarise in \Fig{fig_dyn_spec} our findings for the low-energy spectra of chimera baryons and mesons composed of $({\rm f})$- and $({\rm as})$-type hyperquark constituents. 
To assess the effectiveness of the spectral density method, we further compare our results to those obtained with  generalized eigenvalue problems for the lowest energies and find good agreement. For all three types of chimera baryons, as shown in the figure, the parity-even states are consistently lighter than the odd-parity ones. 
In \Fig{fig_cb_mat_el}, we also present the renormalised overlap factors. For the spin-$1/2$ and parity-even states, which play the role of the top partners in partial compositeness models, $\Sigma_{\rm CB}^+$ turns out to have a larger overlap factor than $\Lambda_{\rm CB}^+$. Our findings are compatible with other gauge theories at the order-of-magnitude level. Further dedicated studies for a wider range of parameter space, and with better control of lattice systematics, will provide quantitative input essential for phenomenological studies.

\section*{Acknowledgement}

E.B. and B.L. are supported by the EPSRC ExCALIBUR programme ExaTEPP project EP/X017168/1. 
E.B. is supported by the STFC Research Software Engineering Fellowship EP/V052489/1. 
E.B., B.L., M.P. and F.Z. are supported by the STFC Consolidated Grant No. ST/X000648/1.
The work of N.F. is supported by the STFC Doctoral Training Grant
No. ST/X508834/1. 
A.L. is funded in part by l'Agence Nationale de la Recherche (ANR), under grant ANR-22-CE31-0011.
D.K.H. is supported by Basic Science Research Program through the National Research Foundation of Korea (NRF) funded by the Ministry of Education (NRF-2017R1D1A1B06033701) and by the NRF grant 2021R1A4A5031460 funded by the Korean government (MSIT). 
L.D.D. and R.C.H. are supported by the STFC grant ST/P000630/1.
L.D.D. is supported by the ExaTEPP project EP/X01696X/1.
J.W.L. is supported by IBS under the project code, IBS-R018-D1. 
H.H. and C.J.D.L. acknowledge support from NSTC Taiwan, through grant number 112-2112-M-A49-021-MY3. 
C.J.D.L. is also supported by the Taiwanese MoST grant 109-2112-M-009-006-MY3. 
B.L. and M.P. are supported by the STFC  Consolidated Grant No. ST/T000813/1.
B.L. was also supported by the STFC Consolidated Grant No. ST/X00063X/1. 
B.L., M.P. and L.D.D. received funding from the European Research Council (ERC) under the European Union's Horizon 2020 research and innovation program under Grant Agreement No.~813942. 
D.V. is supported by STFC under Consolidated Grant No. ST/X000680/1. 
F.Z. is supported by the Advanced ERC grant
ERC-2023-ADG-Project EFT-XYZ.

Numerical simulations have been performed on the Swansea University SUNBIRD cluster (part of the Supercomputing Wales project) and AccelerateAI A100 GPU system (both part funded by the European Regional
Development Fund (ERDF) via Welsh Government), on the local HPC clusters in Pusan National University
(PNU) and in National Yang Ming Chiao Tung University (NYCU).
Numerical simulations have also been performed on the DiRAC Extreme Scaling service Tursa at the University of Edinburgh, and on the DiRAC Data Intensive service DIaL3 at Leicester.
Tursa is operated by the Edinburgh Parallel Computing Centre, and DIaL3 by the University of Leicester Research Computing Service, on behalf of the STFC DiRAC HPC Facility (www.dirac.ac.uk). The DiRAC service at Leicester was funded by BEIS, UKRI and STFC capital funding and STFC operations grants. DiRAC is part of the UKRI Digital Research Infrastructure. 

{\bf Research Data Access Statement}---The results presented in this contribution are published in Ref.~\cite{Bennett:2023mhh,TELOS:2025ash}. 
Analysis workflow and data release are available in Refs.~\cite{data,analysis_dyn,data_dyn}, as per our approach to data reproducibility and open science~\cite{Bennett:2025neg}.

{\bf Open Access Statement}---
The authors have applied a Creative Commons Attribution (CC BY) license to any Author Accepted Manuscript version arising. 


\end{document}